\documentclass[12pt]{iopart}

\usepackage{graphicx}

\usepackage{iopams}

\eqnobysec

\begin{document}
%
\title{Conformal tightness of holographic scaling 
in black hole thermodynamics}

\author{Horacio E. Camblong$^{1}$
and
Carlos R. Ord\'{o}\~{n}ez$^{2}$}

\address{
$^{1}$
Department of Physics, University of San Francisco, San
Francisco, California 94117-1080, USA
\\
$^{2}$ Department of Physics, University of Houston, Houston,
Texas 77204-5506, USA
\\
\vspace*{0.15in}
Email: camblong@usfca.edu}

\vspace*{0.05in}

\begin{abstract}
\newline
The near-horizon conformal symmetry 
of nonextremal black holes
is shown to be a mandatory ingredient for
the holographic scaling of the scalar-field contribution to the black hole entropy.
This {\em conformal tightness\/} is 
revealed by semiclassical
first-principle scaling arguments
through an analysis of the 
multiplicative factors in the entropy due to
 the radial and angular degrees of freedom
associated with a scalar field.
Specifically,
the conformal SO(2,1) invariance of 
the radial degree of freedom
conspires with the area proportionality of the angular momentum sums
to yield a robust holographic outcome.
\end{abstract}


\maketitle

\section{Introduction}
\label{sec:introduction}

The {\em holographic scaling\/} of the Bekenstein-Hawking 
 black hole entropy~\cite{Bekenstein_entropy}: 
\begin{equation}
 S_{\rm BH} \propto {\mathcal A}
 \;  ,
  \label{eq:holo_entropy}
 \end{equation}
  i.e., its proportionality with the area ${\mathcal A}$
of the event horizon,
is one of the central results of contemporary theoretical physics
and a litmus test for quantum theories of gravity~\cite{BH-thermo_reviews}.
In essence, the apparent universality of the {\em area law\/}
 suggests that the horizon encodes information at the quantum 
level~\cite{thooft:85,BH-thermo_reviews,Frolov-Fursaev}
(in the form of a holographic principle~\cite{holographic}).

In the search for patterns that may elucidate this intriguing property,
several lines of research have focused on the existence of a 
near-horizon conformal symmetry~\cite{strominger:98,gov:BH_states,gupta:BH}
as a ubiquitous property of
black hole thermodynamics.
Most remarkably, in Refs.~\cite{carlip:near_horizon,solodukhin:99},
the thermodynamics is explicitly connected with 
the underlying near-horizon conformal field theory
through the Cardy formula.
Alternatively,
within a general framework inspired by 
the brick-wall model~\cite{thooft:85},
conformal quantum mechanics 
(CQM) 
has been directly linked with
the thermodynamic properties~\cite{BH_thermo_CQM,semiclassical_BH_thermo}
 by considering a scalar field
 ($D \geq 4$)
\begin{equation}
S
=
-
\frac{1}{2}
\int
d^{D} x
\,
\sqrt{-g}
\,
\left[
g^{\mu \nu}
\,
\nabla_{\mu} \Phi
\,
\nabla_{\nu} \Phi
+
m^{2} \Phi^{2}
+  \xi R \Phi^{2}
\right]
\; ,
\label{eq:scalar_action}
\end{equation}
 in a given gravitational background.
 Admittedly, in the form derived from the action~(\ref{eq:scalar_action}), 
 it only yields the {\em scalar field contribution\/} to the entropy of the black hole, 
 though extensions to higher-spin fields are in 
progress\footnote{The spin-1/2 case shows similar features; its limiting partition function in spherical coordinates is discussed in Ref.~\cite{PI-fermion_equivalence}.}.
   In addition and most importantly, 
a complete understanding of the predominant role of 
CQM in the 
determination of the thermodynamic properties 
of black holes has not yet been fully achieved,
especially {\em vis-\`{a}-vis\/}
 the physical and geometric meaning of the underlying conformal symmetry. 
It should be noticed that
semiclassical methods,
based on quantum fields in the black hole background, include both the brick wall approach or thermal atmosphere proposal, 
and the entanglement entropy 
approach~\cite{Kabat-Strassler_entanglement,renormalized-entanglement_Cooperman-Luty}.
Moreover, the brick wall
and entanglement entropies are essentially equivalent~\footnote{This 
issue is addressed, for example,
in the review papers of Refs.~\cite{BH-thermo_reviews},
\cite{Frolov-Fursaev},
 and \cite{Soludukhin_entanglement}.}.

Our paper provides further strong 
evidence of the conformal characterization of black hole thermodynamics by 
focusing, for scalar fields,
 on a critical question regarding the entropy, while  
the more complex
issue of the role played by
conformal symmetry in the Hawking effect is left for a forthcoming publication.
In effect,
the apparent central role of CQM 
is highlighted by the conformal determination of the Hawking temperature;
however, 
the angular-momentum sums in semiclassical 
treatments of quantum field theory
also seem to be an essential ingredient in the derivation
of the Bekenstein-Hawking entropy
$ S_{\rm BH} $. 
Thus, the question we now formulate is
{\em whether or not the area law, equation~(\ref{eq:holo_entropy}),
for the entropy is a direct consequence 
of the statistical counting of the angular degrees of freedom\/}, 
independently of the attendant near-horizon conformal symmetry.
In other words,
is the scaling of the entropy critically constrained by the conformal nature of
the radial degree of freedom, above and beyond the
 role played by the angular degrees of freedom? 
Or, by contrast, does the conformal symmetry 
merely provide circumstantial evidence
rather than logical necessity for
the holographic nature of the entropy?
Again, for the sake of simplicity, we will only focus on the scalar-field contribution to the entropy.
This issue is further motivated by early back-of-the-envelope phase-space 
arguments (see~\cite{Susskind-angular_94,Susskind_Lindesay} 
and \cite{quantum-loop-gravity}\footnote{This reference on the loop-quantum gravity approach
suggests a connection with the semiclassical
techniques used in our paper.}),
which appear to suggest the accidental nature of the conformal symmetry,
with $ S_{\rm BH} $ in equation~(\ref{eq:holo_entropy}) arising from the 
straightforward angular summation of degrees of 
freedom.
However,
such arguments rely on implicit 
assumptions regarding the radial degrees of freedom---incidentally,
related observations were made in Ref.~\cite{Marolf_thick-horizon},
where additional issues concerning the nature of the horizon were
also addressed.
Specifically, as we explicitly show below,
the radial part of the metric yields a multiplicative complementary piece in the phase-space counting for the scalar field,
with `radial conformality'
of the near-horizon physics 
guaranteeing the {\em preservation of 
holographic scaling\/}.
What comes out from this analysis is further
evidence of the 
governing role played by CQM---at least for scalar fields:
first, in the direct determination of the Hawking
temperature; and second,
as the central result of our paper,
in the statistical counting of the relevant near-horizon
degrees of freedom. Most importantly, the robustness of
these results can be tested by considering a larger class
of metrics and displaying what may be dubbed 
 {\em conformal tightness\/}:
\begin{quotation}
\noindent
{\em Near-horizon conformal behavior---as described by conformal quantum
mechanics---is a necessary condition for the thermal 
properties of black holes. Any non-conformal
modification of the metric
spoils the thermal identification
of the Hawking temperature
and breaks the holographic scaling or
Bekenstein-Hawking area law for the entropy.}
\end{quotation}

 Our proofs rely on properties of the metric,
via an argument that traces the apparent origin of the 
 holographic scaling of the entropy 
to the angular degrees of freedom. 
A first step is taken in
  section~\ref{sec:area_law}, then
followed by a more precise characterization
  of conformal tightness
in section~\ref{sec:conformal_tightness},
and
  further context in section~\ref{conclusions}.
Additional topics in the appendices include the
computation of the near-horizon spectral integrals 
 (\ref{sec:spectral-integral})
 and alternative 
  phase-space arguments  (\ref{sec:phase-space}).

\section{Origin of the area law:
 basic setup for generalized Schwarzschild metrics} 
\label{sec:area_law}

Our stated goal is to 
highlight the `conformal emergence' of
the area scaling of the black hole entropy,
by considering a quantum scalar field probe
with action~(\ref{eq:scalar_action})
in a gravitational background of the form 
\begin{equation}
 ds^{2}
=
- f (r) \,  dt^{2}
+
\left[ f(r) \right]^{-1} \, dr^{2}
+ r^{2} \,
 d \Omega^{2}_{(D-2)}
\;
\label{eq:RN_metric}
\end{equation}
[in which 
$
d \Omega^{2}_{(D-2)}
$ 
is the metric on the unit $(D-2)$-spheres,
$S^{D-2}$,
that foliate the spacetime manifold
 and
the metric conventions of Ref.~\cite{misner_thorne_wheeler} are used]. 
Equation~(\ref{eq:RN_metric})
represents a static-coordinate description or generalized
Schwarzschild chart for a generic static geometry.
The corresponding family of spacetimes includes the 
Reissner-Nordstr\"{o}m geometries in $D$ spacetime dimensions~\cite{mye:86}, 
and possibly extensions with a cosmological constant.
In Section~\ref{sec:conformal_tightness},
we generalize this family to emphasize
that the conformal nature of the thermodynamics appears to be a fairly 
universal property of black holes.
In principle, the basic
results of {\em conformal tightness\/}
could be further
generalized to axisymmetric stationary spacetimes, as will be discussed
in a forthcoming paper. 
As is well known,
for static spacetimes,
there exists a coordinate frame in which the timelike coordinate $t$
is associated with the Killing vector field
$\mbox{\boldmath  $\xi$ }\!    = \partial_{t}$
that is hypersurface orthogonal~\cite{frolov_novikov}, and 
for which the metric can be adapted to  
$
 ds^{2}
=
g_{00} \,  dt^{2}
+
\gamma_{ij}
 d x^{i} dx^{j}
$,
where
$g_{00}$
and the spatial metric $\gamma_{ij}$ 
are time independent; this includes 
equation~(\ref{eq:RN_metric}) and the generalization
of section~\ref{sec:conformal_tightness}.
The Killing vector
$\mbox{\boldmath  $\xi$ }\! $
permits the introduction of 
a Killing horizon ${\mathcal H}$
where
the surface gravity $\kappa$ 
is defined in purely geometric terms~\cite{frolov_novikov},
\begin{equation}
\kappa
 \stackrel{(\mathcal H)}{=} 
\frac{1}{2}\sqrt{ 
-
\frac{1}{2}
\nabla_{\alpha} \xi_{\beta} 
\nabla^{\alpha} \xi^{\beta} 
   }
\label{eq:surface_gravity_general}
\; .
\end{equation}
In addition,
for the particular class of metrics~(\ref{eq:RN_metric}),
\begin{equation}
\kappa
 \stackrel{(\cal H)}{=}
\sqrt{ \nabla_{\alpha} V (x)  \nabla^{\alpha} V(x)  }
\stackrel{(\mathcal H)}{=} 
\frac{1}{2} \,  f'_{+} 
\, ,
\label{eq:surface_gravity}
\end{equation}
in which 
$V(x) \equiv \sqrt{ - \xi^{\mu} \xi_{\mu}} =
\sqrt{f(r)}$
and all the derivatives
are evaluated at $\mathcal H$,
in terms of
$f_{+}' = f' ( r_{+} )$,
using the standard notation $r=r_{+}$ for the
outer event horizon, identified as the 
largest root of the equation $f(r)=0$.
In the nonextremal case: $f'_{+}  \neq 0$,
which is the focus of this work,
the surface gravity has a well-defined nonzero value
(while the extremal case will be tackled elsewhere, in view
of its subtle analytical properties~\cite{Frolov-Fursaev}).

It should be noticed that the Hawking temperature can be computed
for the metrics~(\ref{eq:RN_metric})
from the removal of the conical singularity,
with the result 
\begin{equation}
\beta =  \frac{ 2 \pi}{\kappa} 
\label{eq:temperature-gravity_connection} 
\end{equation}
for the inverse temperature;
see
section~\ref{sec:conformal_tightness}
for a more detailed account for 
a generalized class of metrics.
As a general methodology,
the relations~(\ref{eq:surface_gravity})
and (\ref{eq:temperature-gravity_connection})---properly
generalized in
section~\ref{sec:conformal_tightness}---will be systematically used
to simplify the interpretation of the statements and calculations;
consequently, our focus will be on
addressing the questions stated in section~\ref{sec:introduction},
especially on the 
conformal nature of the entropy.

 To shed light on the origin of the holographic 
expression~(\ref{eq:holo_entropy})
we will often invoke {\em scaling relationships\/}
arising from the relevant physics,
e.g., 
$\kappa \simeq f_{+}' $.
While  the complete analytical results are displayed
in subsection~\ref{sec:area-law_computation},
the scaling equations
will help elucidate the
role played by the different degrees of freedom 
in the ensuing holographic behavior.

\subsection{Origin of the area law: 
detailed computation
in generalized Schwarzschild coordinates}
\label{sec:area-law_computation}

\begin{figure}[p]
\resizebox{7in}{!}{\includegraphics{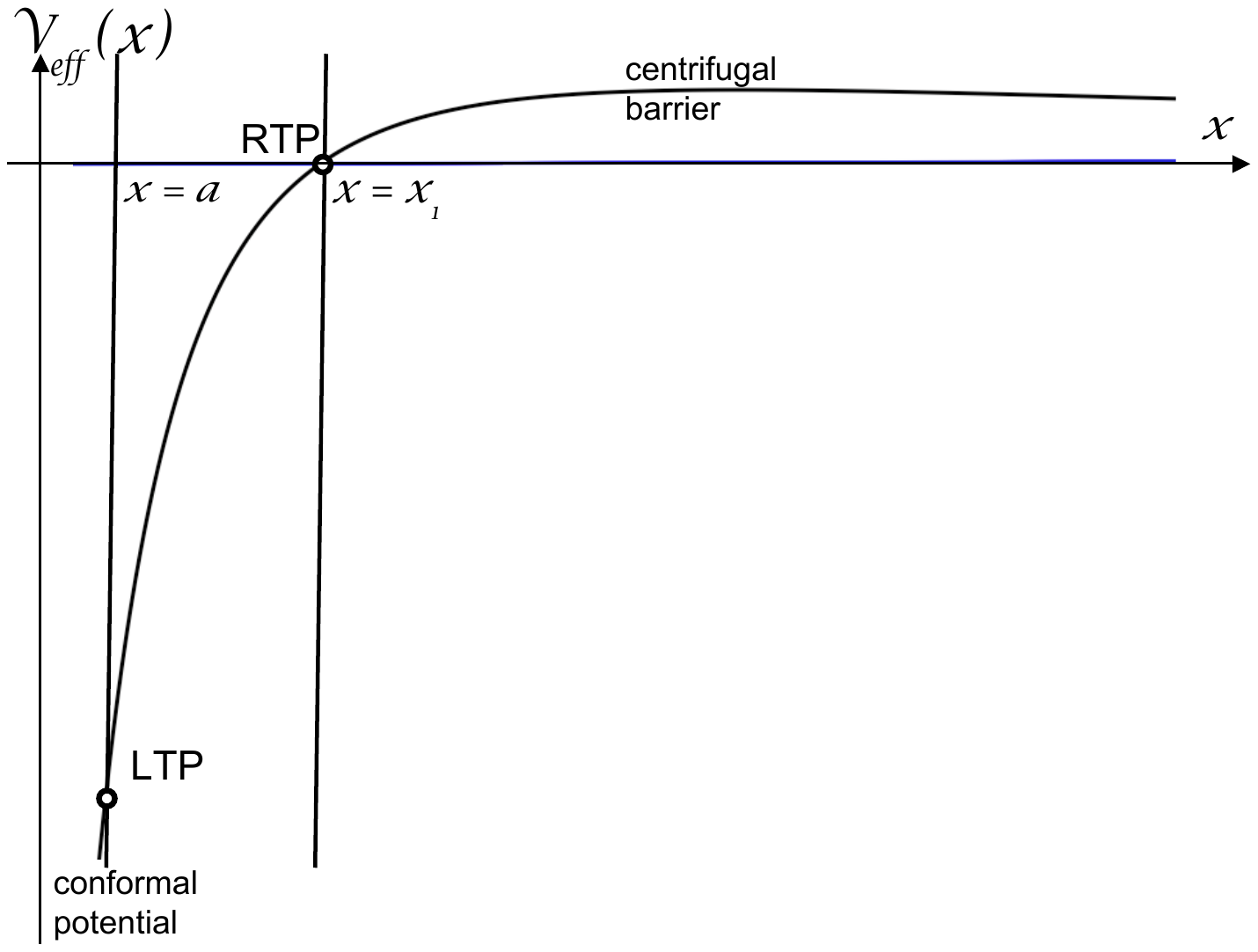}}
\caption{Near-horizon effective gravitational 
potential
$V_{\rm eff}$
from  equation~(\ref{eq:neg-BH-potential}), showing 
the interplay between the angular-momentum and conformal terms.
The WKB left and right  turning points:
$x=a$ and $x=x_{1}$, 
represented by small circles, are labeled
LTP and RTP. }
\label{fig:BH-potential_1}
\end{figure}

The basic strategy is
to identify the origin of the 
area law for the entropy through the bookkeeping
of the relevant degrees of freedom. 
The standard procedure is based on the
Fourier expansion of the quantum field with
a complete set of orthonormal modes
$\Phi_{\omega_{s},s} =
\phi_{s} (r, \Omega )
\,
e^{-i\omega_{s}t}
$
that satisfy the Klein-Gordon equation
in the black-hole background,
\begin{equation}
\frac{1}{ \sqrt{-g} }
\partial_{\mu} \left(
\sqrt{-g} 
\,
g^{\mu \nu}
\,
\partial_{\nu} \Phi
\right)
- \left(
m^{2} 
+ 
\xi R
\right)
\Phi
= 0
\; ,
\label{eq:Klein_Gordon_basic}
\end{equation}
where the terms above include the 
d'Alembert-Beltrami operator, 
 the mass term,
 and a possible  coupling to the Ricci scalar $R$,
 as described by the action~(\ref{eq:scalar_action}).
In this decomposition,
for the static metric~(\ref{eq:RN_metric})
in generalized Schwarzschild coordinates $(t, r,\Omega )$,
the factorization
$\Phi_{\omega_{s}} =
\phi_{s} (r, \Omega )
\,
e^{-i\omega_{s}t}
$
 is based
on the Lie-derivative 
equation 
${\mathcal L}_{
\mbox{\boldmath  ${\scriptstyle \xi}$ }  }
\Phi_{\omega}
 = - i \omega
\Phi_{\omega}
$
with the Killing vector 
$\mbox{\boldmath  $\xi$ }\!   $
(for positive-frequency modes,
and the corresponding conjugate relation for negative-frequency modes).
Thus, 
\begin{equation}
 \Phi(t, r, \Omega   )
= \sum_{s}
\left[
 	a_{s}
\,
\phi_{s} ( r, \Omega   )
\,
e^{-i\omega_{s} t }
 	+ a^{\dagger}_{s}
\,
\phi^{*}_{s} ( r, \Omega   )
\,
e^{i\omega_{s}t}
 	\right]
\; ,
\label{eq:field_Fourier_expansion}
\end{equation}
where the creation and annihilation 
operators $a^{\dagger}_{s}$ and $a_{s}$
satisfy the canonical commutation relations and $s=(n,l,{\bf m})$ labels the radial and angular eigenvalues---in particular, 
the states in the ensuing Fock space are generated according 
to their occupation numbers 
per mode from the Boulware vacuum~\cite{Mukohyama_vac:98}.
Equation~(\ref{eq:Klein_Gordon_basic})
can then be recast in the spatially-reduced form
(with an operator $ \hat{{\mathcal A}}$)
\begin{equation}
\hat{{\mathcal A}}
 \, \Phi  
= 
\omega^{2} \, \Phi
\; ,
\label{eq:Klein-Gordon_curved}
\end{equation}
after the Fourier frequency decomposition above is enforced
(i.e.,
 $\partial_{t}^{2} \rightarrow - \omega^{2}$).
 The spatial part of each mode is further rewritten
in Liouville normal form~\cite{BH_thermo_CQM}
by the substitution
$
\phi 
(r,\Omega )
= 
Y_{l {\bf m}  } \big(  \Omega \big)
\,
u (r)
\,
[f(r)]^{-1/2} 
\,
r^{-(D-2)/2}
$,
where 
the angular dependence of the ultraspherical harmonics
$Y_{l {\bf m} } \big( \Omega \big)$
(with eigenvalues  $\lambda_{l,D} 
=l(l+D-3) $)\footnote{The separation of the angular degrees of freedom is defined more precisely
in subsection~\ref{sec:thermo-landscape}.} 
supplements the normal radial behavior
\begin{equation}
u''(r) 
- V_{\rm eff}  (r; \omega, \alpha_{l,D}  ) 
\,
u (r) 
=0
\;  .
\label{eq:Klein_Gordon_normal_radial}
\end{equation} 
In equation~(\ref{eq:Klein_Gordon_normal_radial}),
 the primes stand for radial derivatives
and
$V_{\rm eff} (r)$
is interpreted as an effective gravitational potential.
Defining 
\begin{equation}
\alpha_{l,D} 
=
l(l+D-3) 
+ \nu^{2} = 
\left(
l + \frac{D-3}{2} 
\right)^{2}
\; ,
\label{eq:ang-momentum_coupling}
\end{equation}
with
$
\nu = (D-3)/2$, which
yields the angular momentum eigenvalues for each dimensionality $D$,
 and
$\Theta= \omega/2 \kappa $,
then, the {\em dominant near-horizon
orders of the effective interaction\/}
are~\footnote{These results are generalized in subsection~\ref{sec:thermo-landscape},
where the steps leading to Eqs.~(\ref{eq:Klein_Gordon_normal_radial})
and
(\ref{eq:neg-BH-potential}) 
are made explicit.}
\begin{equation}
V_{\rm eff}  (r; \omega, \alpha_{l,D}  ) 
\stackrel{(\mathcal H)}{\sim}
-
\left(
\Theta^{2}
+
 \frac{1}{4} 
\right)
\frac{1}{ x^{2} }
+
\frac{ \alpha_{l,D}  }{2 \kappa \,  r_{+}^{2}}
\frac{1}{x}
\; 
\label{eq:neg-BH-potential}
\end{equation}
[up to terms $O(x)$],
where 
$x= r -r_{+}$
and 
hereafter we use the symbol
$ \stackrel{(\mathcal H)}{\sim} $
for the near-horizon expansion
with respect to the variable $x$,
which displays the typical pattern
shown in Figure~\ref{fig:BH-potential_1}~\footnote{This is not the usual effective potential 
found in textbooks~\cite{misner_thorne_wheeler},
which, unlike our choice of Eqs.~(\ref{eq:Klein_Gordon_normal_radial})
and (\ref{eq:neg-BH-potential}), typically involves the tortoise coordinate.}.
In particular,
the SO(2,1) conformal 
symmetry~\cite{jackiw&alfaro} 
under scrutiny is displayed by the leading 
term
in equation~(\ref{eq:neg-BH-potential}),
which amounts to 
the scale invariant potential
$V_{\rm eff} 
 \stackrel{(\mathcal H)}{\sim} 
\left. 
V_{\rm eff} 
\right|_{\rm CQM}
=
-
\left( \Theta^{2} + 1/4 \right)/x^{2}
$, 
i.e., it
has same degree of homogeneity as the
derivative term in equation~(\ref{eq:Klein_Gordon_normal_radial}). 
The `conformal parameter'
$
\Theta
$
characterizes the effective coupling strength of
CQM~\cite{anomaly_qm_ISP_anyD,renormalization_CQM}
and determines the Hawking temperature~\cite{Hawking_temperature} 
in a unique manner~\cite{padmanabhan_HawkingT}.
However, the explicit form of the 
effective gravitational 
potential in equation~(\ref{eq:neg-BH-potential})
also reveals that the naive power counting 
with respect to $1/x$ breaks down due to the
competing interplay
between the two leading terms in the near-horizon expansion:
\begin{quotation}
\noindent
(i)
The purely conformal potential
$\left. V_{\rm eff} \right|_{\rm CQM} \propto - 1/x^{2}$,
which carries the radial degrees of freedom,
 is strictly dominant 
for low angular momenta
$\alpha
\equiv
\alpha_{l,D} $.

\noindent
(ii) 
The centrifugal potential term
 $ \alpha/\left( f r^{2} \right)
\propto 1/x$,
which carries the angular degrees of freedom, 
becomes comparable to the conformal interaction
for high angular momenta $\alpha$.
\end{quotation}
Evidently,
this {\em interplay is the key\/} to answering the
questions posed in section~\ref{sec:introduction},
{\em regarding the relative roles played
by the angular and radial degrees of freedom.\/}

With the above considerations in mind, we seek to reformulate the
thermodynamics within this framework, in purely conformal terms.
In this program, the critical components  
of the entropy calculation involve the sequential
computation of the following
quantities: the spectral function, the entropy,
and Planck-scale renormalization.

First,
the computation of the spectral 
function $N(\omega)$ from the near-horizon effective gravitational 
potential~(\ref{eq:neg-BH-potential})
can be accomplished either
by WKB or phase-space methods~\cite{BH_thermo_CQM,semiclassical_BH_thermo},
and also via the heat-kernel approach~\cite{holographic_scaling_I}.
For our presentation in this section,
the competition between the angular-momentum degrees
of freedom and the leading near-horizon (conformal) effective 
potential
is most clearly seen in its one-dimensional WKB form, 
which furnishes the asymptotically exact `spectral integral'
\begin{equation}
\! \! \! \! \! \! \! \! \! \! \! \!
 N(\omega) 
\stackrel{(\mathcal H)}{\sim}
\frac{ \Theta}{\pi \, \Gamma (D-2) }
\int_{0}^{ \alpha_{\rm max} (a) } 
d \alpha \,
\alpha^{D/2- 2} 
 \, 
\int_{a}^{ x_{1} (\alpha) } 
\frac{dx}{x}
\,
\sqrt{ 1 
- 
\frac{ \alpha }{2 \kappa r_{+}^{2} \, \Theta^{2} }
\, x }
\; 
\label{eq:WKB_number_of_states}
\end{equation}
as the near-horizon limit of the 
WKB algorithm~(\ref{eq:semiclassical_number_of_states})
further discussed in~\ref{sec:spectral-integral}.
In equation~(\ref{eq:WKB_number_of_states}),
$a$ is a coordinate cutoff defining the 
left turning point
of the potential~(\ref{eq:neg-BH-potential})
with `zero effective energy,'
while
$x_{1} = r_{1} - r_{+} $
is the root of the radicand associated with its
right turning point:
$x_{1} (\alpha)
=
2 \kappa r_{+}^{2} \, \Theta^{2} /\alpha$
and the
angular momentum upper bound
$\alpha_{\rm max} (a)
=
2 \kappa r_{+}^{2} \, \Theta^{2}/a
$
corresponds to the 
left-right
turning-point
orientation
 $a \leq x_{1} (\alpha)$.
An alternative phase-derivation is presented in~\ref{sec:phase-space}.
The `conformal measure'
$dx/x$ of the radial integration
in equation~(\ref{eq:WKB_number_of_states})
plays a key role in our derivation and is related
to the scale-invariant radial measure
used in Ref.~\cite{Marolf_thick-horizon}.

Second,
the computation
of the entropy itself can be performed
for the field~(\ref{eq:scalar_action})
with the canonical-ensemble
spectral rule~\cite{BH_thermo_CQM,semiclassical_BH_thermo}
\begin{equation}
S
=
- \int_{0}^{\infty}
d \omega  
\,
\ln (1 - e^{-\beta \omega})
\,
\left[
\left(
\omega \frac{d}{d \omega} + 2
\right)
\frac{dN(\omega)}{d\omega}
\right]
\; ,
\label{eq:entropy_formula}
\end{equation}
which amounts to a thermal free-field distribution
of independent modes at inverse temperature $\beta$,
with canonical partition function
$
Z=
\exp \left[
 - \int_{0}^{\infty}
d N ( \omega)  
\,
\ln (1 - e^{-\beta \omega})
\right]
$.
The details of the spacetime curvature background, 
carried by the effective gravitational potential,
are embodied in the nontrivial spectral function $N(\omega)$
of equation~(\ref{eq:WKB_number_of_states}).

As a final step,
the evident near-horizon divergence
$
N (\omega)
 \stackrel{(\mathcal H)}{\propto} 
a^{-(D/2-1)}
$
of equation~(\ref{eq:WKB_number_of_states}).
calls for regularization,
e.g., 
in terms of a short-scale coordinate regulator $a$
to be introduced as the lower limit in 
the divergent integral 
$\int_{a} dx $.
The ensuing `ultraviolet catastrophe' 
of the spectral number function    
$N(\omega) $
can be viewed as arising from
the ultraviolet singular nature of CQM.
As first proposed in Ref.~\cite{thooft:85},
for the metrics~(\ref{eq:RN_metric}),
a geometric radial distance from the horizon,
 \begin{equation}
h_{D} =
\int_{ r_{+}}^{r_{+}+a} ds
\stackrel{(\mathcal H)}{\sim}
\sqrt{ \frac{2 a}{\kappa  } }
\; ,
\label{eq:brick_wall_geometrical_elevation_regularized}
\end{equation}
yields an invariant answer that reveals new spacetime physics 
at the Planck scale $\ell_{P} \simeq h_{D}$.

The three essential steps listed above will now be used to show 
the {\em conformal tightness\/} of the black hole entropy.
Equation~(\ref{eq:WKB_number_of_states})
can be evaluated 
as shown in~\ref{sec:spectral-integral},
leading to 
the spectral function
 \begin{equation}
 \! \! \! \! \! \!
 \! \! \! \! \! \! \! \! \! \! \! \!
 \! \! \! \! \! \! \! \! \! \! \! \!
 N(\omega)
\stackrel{(\mathcal H)}{\sim}
\frac{2}{\pi
\,   (D-2) \, \Gamma (D-2) }
\,
B \left( \frac{D-2}{2} , \frac{3}{2} \right) \,
\Theta^{D-1} 
\,
\left( 2 \kappa \, r_{+}^{2}  \right)^{ (D-2)/2 }
\,
a^{-(D/2- 1)}
\; ,
\label{eq:WKB_number_of_states_2}
\end{equation}
where
$B(p,q)$ is the beta function.
Consequently,
after the 
replacement~(\ref{eq:brick_wall_geometrical_elevation_regularized})
is made,
the relevant spectral-function scaling becomes
\begin{equation}
 N(\omega)
\simeq
\biggl(
\frac{\omega}{  2 \kappa} \biggr)^{D-1}
\;
\left(
\frac{
{\mathcal A}_{D-2} }{ \left[ h_{D} \right]^{D-2} }
\right)
\; ,
\label{eq:WKB_number_of_states_scaled}
\end{equation}
by the explicit introduction of the area 
${\mathcal A}_{D-2} $ from factors $r_{+}^{D-2}$;
the ratios highlight the dimensionless nature of
$ N(\omega)$.
The numerical proportionality prefactor
in 
equation~(\ref{eq:WKB_number_of_states_scaled}),
as needed for a replacement of the symbol
$\simeq$ by an equal sign,
can be written as
\begin{equation}
\widehat{\mathcal N}_{D}
=
\frac{ 2^{D-1} \, B \left( (D-2)/2 , 3/2 \right) }{
\pi \,   (D-2) \, \Gamma (D-2) \, \Omega_{D-2} }
\; .
\label{eq:WKB_number_of_states_prefactor}
\end{equation}

Finally,
the entropy~(\ref{eq:entropy_formula})
corresponding to Eqs.~(\ref{eq:WKB_number_of_states_scaled})
and (\ref{eq:WKB_number_of_states_prefactor})
is given by
\begin{equation}
S
\stackrel{(\mathcal H)}{\sim}
{\mathcal S}_{D}
\,
S_{\rm BH}
\,
\left(
\frac{ 2 \pi}{ \kappa \beta}
\right)^{D-1}
\; ,
\label{eq:entropy_brick_wall}
\end{equation}
which 
consists of three distinct factors:
the numerical prefactor 
${\mathcal S}_{D}$,
the Bekenstein-Hawking entropy
$S_{\rm BH}
=
\,
{\mathcal A}_{D-2}/4 \left( \ell_{D} \right)^{D-2}  
$
and the temperature factor
$\left( 2 \pi/ 
 \kappa \beta  \right)^{D-1}$.
While the latter 
can be straightforwardly
set equal to unity due the Hawking inverse-temperature
identification~(\ref{eq:temperature-gravity_connection}),
the former
involves some additional algebra~\cite{BH_thermo_CQM}
in terms of the Riemann zeta function $\zeta (s)$,
leading to
$
{\mathcal S}_{D} 
\left[ h_{D}/\ell_{D} \right]^{(D-2)}
= 
  D \zeta (D) \Gamma (D/2-1)  \pi^{1-3D/2}/2^{D-2}
$,
which shows ${\mathcal S}_{D} $ is a number of order unity,
if and only if the elevation $h_{D}$ 
is of the order of the Planck length $\ell_{D}$.
Thus, for the 't Hooft assignment~\cite{thooft:85}
(generalized to $D$ spacetime dimensions)
\begin{equation}
 h_{D}/\ell_{D} 
=
\frac{1}{2}
\left[
 D \zeta (D) \Gamma (D/2-1) \pi^{1-3D/2}
\right]^{1/(D-2)}
\; ,
\label{eq:brick_wall_geometrical_elevation}
\end{equation}
which is a form of geometric renormalization,
the entropy reduces to the holographic result 
$S_{\rm BH}$.

\subsection{Origin of the area law: 
scaling arguments
in generalized Schwarzschild coordinates}
\label{sec:area-law_scaling}

In this subsection,
we provide further insight into
the {\em scaling\/}
relations that 
constitute the core of the 
holographic scaling of the entropy.
Unlike the results of the previous subsection,
we will omit numerical factors,
assume naturalness, and enforce dimensional relations
defined up to 
dimensionless factors of order one.

First,
 the value of the Hawking temperature 
$T_{H} \equiv \left( \beta_{H} \right)^{-1}$,
in addition to its connection with the surface gravity via
equation~(\ref{eq:temperature-gravity_connection}),
implies the temperature--inverse-length chain of scaling relations
\begin{equation} 
2 \pi \left( \beta_{H} \right)^{-1}
\simeq
\kappa 
\simeq
f_{+}'
\; ,
\label{eq:temp-curvature_scaling-chain}
\end{equation}
 where $f'_{+}$ is the natural scale of inverse length arising from a nonextremal metric~(\ref{eq:RN_metric})
via the near-horizon expansion $f \sim f'_{+} x$. 
The singular case of extremal black holes involves additional subtleties~\cite{nonextremal-BH-entropy_refs}
 and will be discussed elsewhere\footnote{It should be noticed
that the scaling chain of equation~(\ref{eq:temp-curvature_scaling-chain}) is 
valid for {\em any nonextremal black hole\/}.
In addition,
$ f_{+}'  \simeq 1/r_{+} \simeq (2 G_{N}M)^{-1}$,
where $M$ is the black-hole mass and $G_{N}$ is Newton's gravitational constant,
is only 
a rough estimate for black holes sufficiently away from extremality.}.
 Hereafter, the symbol 
 $\simeq $
  stands for a binary scaling relation 
 in the sense of dimensional analysis,
 governed by
the following criteria:
(i)  completeness of relevant scale dependence;
(ii) 
a `naturalness' condition satisfied
by its numerical coefficients
 (i.e., with their 
order of magnitude being
not too different
 from order one).
Such dimensional equality
 captures the scale homogeneity of any quantity with respect to all other
{\em relevant physical or geometrical parameters\/} 
for black hole thermodynamics.

The one additional ingredient not
displayed in equation~(\ref{eq:temp-curvature_scaling-chain})
is the spectral scaling 
with respect to the inverse temperature $\beta$,
i.e., the dimensional relationship between 
$\beta$ and a {\em characteristic frequency \/} 
$\omega$.
For the thermodynamic functions,
this scaling 
is driven by
the thermal distribution,
e.g., as in
 equation~(\ref{eq:entropy_formula})
for the entropy.
Thus, the temperature dependence is 
fully controlled by the 
spectral partition function 
$L(\xi) =
-
\ln (1 - e^{-\xi} )$,
through the Boltzmann factor
 $e^{-\xi} $.
As this dependence is 
carried by a function
of $\xi = \beta \, \omega$ 
alone,
the characteristic frequency 
is given by the dimensional equality
\begin{equation} 
\omega_{\rm characteristic}
\simeq
\beta^{-1}
\; ,
\label{eq:temp-frequency_relation}
\end{equation}
which further supplements
the chain of scaling relations~(\ref{eq:temp-curvature_scaling-chain}).
This 
can be seen from the following 
fairly general argument for a generic thermodynamic function
\begin{equation}
{\mathcal T}
=
 \int_{0}^{\infty}
d \omega
\,
L (\beta \omega)
\,
\hat{\mathcal T}_{\omega} (\beta)
\!
\left[
\frac{ dN(\omega) }{ d\omega }
\right]
\; ,
\label{eq:generic_thermo-function}
\end{equation}
where
the operator 
$\hat{\mathcal T}_{\omega}(\beta)$
governs the specific properties of the
thermodynamic function
${\mathcal T}$.
In general, $\hat{\mathcal T}_{\omega} (\beta)$
is a differential operator with respect to $\omega$ and possibly 
$\beta$-dependent,
as follows by the replacement of any
thermal derivatives 
$\partial/\partial \beta = 
\omega \partial/\partial \xi $
and
integration by parts
with respect to $\xi$; this effectively replaces all 
such derivatives by their frequency counterparts,
thus yielding the operator 
$\hat{\mathcal T}_{\omega}(\beta)$.
For example,
 for the free energy ${\mathcal T} = F$, 
$\hat{\mathcal T}_{\omega}= - 1/\beta$;
and for the entropy
${\mathcal T} = S$, 
$\hat{\mathcal T}_{\omega}=   2 + \omega \, d/d \omega$.
As a result, it is
the interplay of the three factors in
equation~(\ref{eq:generic_thermo-function})---with the third 
one being the density of states---that yields the 
corresponding thermodynamic potential function.
The scaling statement~(\ref{eq:temp-frequency_relation})
follows by enforcing the replacement $\xi = \beta \omega$,
with $\xi \simeq 1$ being the dominant part of the 
integral---this procedure is elaborated upon below 
for the all-important entropy function.

In particular,
the characteristic dimensional
scaling of the entropy~(\ref{eq:entropy_formula})
can be seen
by rewriting it in the form
\begin{equation}
S
=
- \int_{0}^{\infty}
d \xi
\,
\ln (1 - e^{-\xi})
\,
\left[
\left(
\xi \frac{d}{d \xi} + 2
\right)
\frac{dN(\xi/\beta)}{d\xi}
\right]
\; ,
\label{eq:entropy_formula_dimensionless}
\end{equation}
and recognizing that the dominant value of the integral 
arises from $\xi \simeq 1$
as the interplay between the decreasing logarithmic partition function and the increasing density of states. 
As all the factors in 
equation~(\ref{eq:entropy_formula_dimensionless})
are dimensionless and of order one, except for 
the spectral function $N(\omega)$,
it follows that the final outcome is a number
governed by naturalness times the spectral function itself
evaluated at $\xi/\beta \simeq 1/\beta$.
This argument is unambiguous when the functional form of 
$N(\omega)$ is homogeneous, which is certainly the case for our problem, e.g., 
as described in equation~(\ref{eq:WKB_number_of_states_scaled}); 
this power-law dependence $\omega^{p}$ with $p=D-1$ implies,
from Euler's theorem for homogeneous functions, that the integral of equation~(\ref{eq:entropy_formula_dimensionless})
scales as $\beta^{-p}$, which amounts to 
the shorthand expression
\begin{equation}
S
\left.
\simeq
N(\omega) \right|_{\omega \simeq \beta^{-1}}
\; .
\label{eq:entropy_number-density_scaling}
\end{equation}
Therefore, for any homogeneous density function, the scaling~(\ref{eq:entropy_number-density_scaling}), 
with the interplay between the partition-like function and the density of states, involves a numerical coefficient of order one.
However, the details are subtle, in the presence of a regularization,  and are further 
discussed in the next section.

Consequently,
the scaling~(\ref{eq:WKB_number_of_states_scaled})
is directly transferred from the spectral function 
to the entropy, with the additional
simplification that
the Hawking-temperature identification is made.
Once this is recognized, 
the entropy acquires the factorized structure
\begin{equation}
 S 
\simeq
\underbrace{
\left(
\begin{array}{c}
{\rm `Extra}
\\
{\rm Scaling}
\\
{\rm Factor'}
\end{array}
\right)
}_{ \mbox{\small RADIAL}}
 \;
 \times
 \;
 \underbrace{
 \begin{array}{c}
\\
\; \; \; 
{\mathcal A} 
\; \; \; .
\\
\\
\end{array}
 }_{ \mbox{\small ANGULAR}}
 \label{eq:entropy_angular-radial-components}
\end{equation}
In short, this factorization for $S$ follows from~(\ref{eq:entropy_number-density_scaling})
from the corresponding factorization of 
$N(\omega)$, with an area factor $\mathcal{A}$, e.g., equation~(\ref{eq:entropy_number-density_scaling}).
This is a direct consequence of the statistical counting and is most easily seen via the phase space methods, as in appendix B. The phase counting~(\ref{eq:phase_space_spectral_number_explicit_fully_multidim})
  shows directly, for the thermal atmosphere of a black hole, 
that $N \propto \mathcal{A} \int dr ||\vec{k}||^{d}$, 
because it has an entropy spread over a shell of
volume $dV_{(d)} \sim \Omega_{(d-1)} r_{+}^{d-1}dr \sim \mathcal{A} \, dr$;
and the integral depends on the details of the radial problem, thus yielding an `extra scaling factor'.

The structural simplicity of this result provides the 
natural context to address 
{\em the relative roles played
by the angular and radial degrees of freedom\/}
for the area law of the entropy.
Moreover,
for the general-relativistic black holes described by the 
standard metric~(\ref{eq:RN_metric}):
(i)
the near-horizon radial physics is conformal;
and (ii) 
the ensuing scale-invariant nature of the radial degree of freedom 
implies the {\em trivial scaling\/}
\begin{equation}
\left.
\left(
\begin{array}{c}
{\rm `Extra}
\\
{\rm Scaling}
\\
{\rm Factor'}
\end{array}
\right)
\right|_{{\rm CQM}; \; \beta \equiv \beta_{H} }
\simeq
1
 \; ,
 \label{eq:angular-radial_entropy-resolution}
\end{equation}
which obviously generates the area law in
equation~(\ref{eq:entropy_angular-radial-components}).
Therefore, 
equation~(\ref{eq:angular-radial_entropy-resolution}),
which is thus traced back to CQM,
involves a preservation of holographic scaling---the radial
degree of freedom does not spoil
the area law suggested by 
equation~(\ref{eq:entropy_angular-radial-components}).

In conclusion, our procedure shows that
the separation of degrees of freedom leads to the 
factorized form of the entropy~(\ref{eq:entropy_angular-radial-components}),
which, in turn,
generates the Bekenstein-Hawking area law via a trivial 
radial scaling.
Insofar as equation~(\ref{eq:angular-radial_entropy-resolution})
relies on the scale-invariant 
nature of the near-horizon radial degree of freedom, 
this result can be attributed to
the role played by CQM. 
A more exhaustive characterization 
is given in the next section.

\section{Conformal tightness of black hole thermodynamics: 
a general derivation}
\label{sec:conformal_tightness}

In this section we examine
the anticipated {\em conformal tightness\/} of the area law
by specifying the generalized conditions for
Eqs.~(\ref{eq:entropy_angular-radial-components})
and (\ref{eq:angular-radial_entropy-resolution})
to hold true.
For this purpose,
we simply relax the form of equation~(\ref{eq:RN_metric});
as a first step, we consider
a generic spherically symmetric metric,
with extensions for axisymmetric metrics
to be discussed elsewhere.

\subsection{Thermodynamic landscape of black hole solutions}
\label{sec:thermo-landscape}

For the analysis of black hole solutions
{\em vis-\`{a}-vis\/}
thermodynamic and conformal properties,
we simply relax the form of equation~(\ref{eq:RN_metric})
into
\begin{equation}
 ds^{2}
=
- A (r) \,  dt^{2}
+
B (r) \, dr^{2}
+ r^{2} \,
 d \Omega^{2}_{(D-2)}
\; , 
\label{eq:off-shell_metric}
\end{equation}
which is the most general spherically symmetric metric.
This geometry
is described via
a further generalization of Schwarzschild coordinates, i.e.,
with a time $t$ measured by an observer at
asymptotic infinity
and with
$r$ being an  `area radial coordinate'
[such that a  $(D-2)$-sphere 
of radius r has proper area
${\mathcal A} = \Omega_{D-2} \, r^{D-2}$].
As such, the new family
encompasses a larger 
class of black hole metrics than
equation~(\ref{eq:RN_metric})---as in the stringy example of 
subsection~\ref{sec:further-CT_examples};
in addition, it may be regarded
as a generic  `off-shell metric'
(not necessarily a solution to the general-relativistic field equations)
conceived to characterize
the conformal tightness property.
Within the family of geometries described by 
equation~(\ref{eq:off-shell_metric}),
one may consider the hypersurfaces with  $r= $ constant 
and decrease $r$ gradually from asymptotic spacelike infinity,
until the value $r=r_{+}$ is reached for which the 
hypersurface is everywhere null---this is a boundary 
where ingoing timelike paths cannot go back to infinity;
thus, it is an event horizon ${\mathcal H}$.
As $\partial_{\mu} r$ is normal to these
hypersurfaces, the inverse metric element $g^{rr} =
g^{\mu \nu}  \partial_{\mu} r 
\partial_{\nu} r $ vanishes, so that 
$
 g^{rr} (r_{+})= 0$
and
\begin{equation}
\left.
B (r) 
\right|_{\mathcal H} = \infty
\; .
\label{eq:roots_g^rr}
\end{equation}
Therefore, in this approach, one identifies the roots of 
$g^{rr}(r) = B^{-1}(r)$
and selects the largest one,
defining ${\mathcal H}$ and leading to the analysis
of the field behavior in its neighborhood.

For the concomitant  black hole thermodynamics, we now
extend the statistical-counting treatment 
of subsection~\ref{sec:area-law_scaling}
with the generalized background~(\ref{eq:off-shell_metric}).
The statistical thermodynamic details are computed in subsection~\ref{sec:spectral-density_CT}
after all the relevant quantities and parameters are reevaluated.
For the generalized metrics~(\ref{eq:off-shell_metric}),
the Klein-Gordon equation~(\ref{eq:Klein_Gordon_basic}) has the 
 explicit form
  \begin{eqnarray}
& & 
\! \! \! \! \! \! \! \! \! \! \! \!
\! \! \! \! \! \! \! \! \! \! \! \!
\! \! \! \! \! \! \! \! 
\left[ 
\Box 
- 
\left(
m^{2} 
+ 
\xi R
\right)
 \right] 
 \Phi
\nonumber
\\
&   & 
\! \! \! \! \! \! \! \! \! \! \! \!
\! \! \! \! \! \! \! \! \! \! \! \!
\! \! \! \! \! \! \! \! 
= \left\{
- \frac{1}{ A }
\,
\partial_{t}^{2}
+
\frac{1}{ B }
\,
\left[
\partial_{r}^{2}
+
\frac{1}{ \hat{f} r^{D-2}}
\partial_{r}
\left(
\hat{f} r^{D-2} 
\,
\partial_{r}
\right)
\right]
+ 
\frac{1}{r^{2} }
\,  
{\Delta}^{(D-2)}_{(\sigma)} 
- 
\left(
m^{2} 
+ 
\xi R
\right)
\right\}
\Phi
=
0
\; ,
\label{eq:Klein-Gordon_curved-0_with-L}
\end{eqnarray}
where a generalized scale factor 
\begin{equation}
\hat{f} \equiv \hat{f} (r) = \sqrt{ \frac{A}{B} }
\label{eq:GM_scale-factor}
\end{equation}
 is defined,
and
 ${\Delta}^{(D-2)}_{(\sigma)} $
is  the Laplacian on $S^{D-2}$.
 The angular dependence is given by 
 the spherical harmonics
$Y_{l{\bf m}} \big( \Omega \big)$,
i.e., the  eigenfunctions of 
 the negative Laplacian $
-{\Delta}^{(D-2)}_{(\sigma)} $
on  $S^{D-2}$,
\begin{equation}
 -\frac{1}{\sqrt{\sigma}}
 	\frac{\partial}{\partial \theta^{a}}
\left[
\sqrt{\sigma}
\,
\sigma^{ab}
 	\frac{\partial}{\partial \theta^{b}}
Y_{l{\bf m}} \big( \Omega \big)
\right] 
=
l(l+D-3) 
Y_{l{\bf m}} \big(  \Omega \big)
\label{eq:spherical_harmonics}
\; ,
\label{eq:angular_eigenvalues}
\end{equation}
for an arbitrary choice of angular coordinates
$\Omega \equiv \left\{ \theta^{a} \right\} $ ($a=1, \ldots , D-2$) and
metric
$\sigma_{ab}$ on  $S^{D-2}$.
The assumed spherical symmetry of the gravitational 
background~(\ref{eq:RN_metric})
implies that both $\alpha_{l,D}$ and 
$\omega = \omega_{nl}$ are independent of the
angular momentum quantum numbers ${\bf m}$.
In particular, the multiplicity $g_{l}$ of the 
eigenvalue in
equation~(\ref{eq:spherical_harmonics}) 
is
$
g_{l} = (2l+D-3) (l+ D-4)!/[l! (D-3)!]
$.
Then, a complete set of 
orthonormal solutions
reads
\begin{equation}
\phi_{nl{\bf m}} (r,\Omega )
= 
Y_{l{\bf m}} \big(  \Omega \big)
\chi (r)
u_{nl} (r)
\; ,
\end{equation}
where the Liouville transformation factor
$\chi (r) = \left[ \hat{f} (r) r^{D-2}\right]^{-1/2}$
is used 
to reduce the 
 radial equation
to its Liouville-normal  form,
\begin{equation}
u_{nl}''(r) 
+
I_{(D)} (r; \omega_{nl}, \alpha_{l,D}  ) 
\,
u_{nl}(r) 
=0
\;  ,
\label{eq:Klein_Gordon_normal_radial2}
\end{equation}
for every  particular frequency
$\omega_{nl}$,
with
\begin{eqnarray}
\! \! \! \! \!  \! \! \! \! \! 
\! \! \! \! \!  \! \! \! \! \! 
\! \! \! \! \!  \! \! \! \! \! 
I_{(D)}   
(r; \omega, \alpha_{l,D}  ) 
 & \equiv &  
-V_{\rm eff}  (r; \omega, \alpha_{l,D}  ) 
\nonumber
\\
 \! \! \! \! \!  \! \! \! \! \! 
\! \! \! \! \!  \! \! \! \! \! 
\! \! \! \! \!  \! \! \! \! \! 
& = &
\frac{ 1 }{\hat{f}^{2}}
\left(
 \omega^{2} + 
\frac{\hat{f}'^{2}}{4}
\right)
-
B \,
\frac{\alpha_{l,D} }{ r^{2} }
\nonumber
\\
& & 
-B \, \left(  m^{2} + \xi R \right)
-
\frac{\hat{f}''}{2 \hat{f} }
-
\frac{(D-2)}{r} \, \frac{ \hat{f}'}{2  \hat{f}}  
+
\left[
\left(
B - 1
\right)
\nu^{2} 
+ \frac{1}{4}
\right]
\frac{1}{r^{2}} 
\;  ,
\label{eq:Klein_Gordon_normal_radial_invariant}
\end{eqnarray}
where
the angular-momentum coupling $\alpha_{l,D} $ is defined 
in equation~(\ref{eq:ang-momentum_coupling}).

Equation~(\ref{eq:Klein_Gordon_normal_radial_invariant})
is the basis  for our analysis for the remainder of this paper.
 To determine the relevant contributions to black hole thermodynamics,
 we perform a near-horizon expansion,
 assuming that the metric coefficients 
 admit the leading expressions
 \begin{eqnarray}
 A (r)
&  \stackrel{(\mathcal H)}{\sim} &
 c_{t} x^{p_{t}}
  \;  ,
\label{eq:A-exp_off-shell}
\\
\left[ B(r) \right]^{-1}
&   \stackrel{(\mathcal H)}{\sim}  &
 c_{r} x^{p_{r}}
\label{eq:B-exp_off-shell}
  \; ,
  \end{eqnarray}
where $c_{t}, c_{r} > 0$
from generic metric-signature
conditions,
 while 
\begin{equation}
p_{r} > 0
\label{eq:p-r_horizon-condition}
\end{equation}
 conforms to the existence of a horizon via equation~(\ref{eq:roots_g^rr}).
Notice, in particular, that
\begin{equation}
\hat{f}  
\stackrel{(\mathcal H)}{\sim}
\sqrt{c_{t} c_{r} }
\;
 x^{p/2}
\; ,
\end{equation}
where
\begin{equation}
 p
 =
  p_{t}  + p_{r} 
  \; .
\label{eq:nh-potential-exponent_off-shell}
 \end{equation}

As a simple power-counting in equation~(\ref{eq:Klein_Gordon_normal_radial_invariant}) shows,
when the near-horizon expansion  is enforced, 
 the only relevant terms become the first two in equation~(\ref{eq:Klein_Gordon_normal_radial_invariant}),
 which provide the
leading interaction (first one)
and the angular momentum contribution (second one).
In this approach, $p_{t}$ and $p_{r}$
may be regarded as free parameters to explore
{\em the `thermodynamic landscape'
of possible black holes\/};
as it turns out, the following analysis
inevitably brings us back to CQM.
 The main results for the relevant functions and parameters
are outlined in the next few subsections.

\subsection{Modified effective potential and angular momentum}
\label{sec:eff-potential_CT}

The leading near-horizon gravitational potential
arises from the
first term in the second line of equation~(\ref{eq:Klein_Gordon_normal_radial_invariant}), and 
takes the form
 \begin{equation}
 V_{\rm eff} (x) 
 \stackrel{(\mathcal H)}{\sim} 
 - 
 \frac{ 1 }{\hat{f}^{2}}
\left(
 \omega^{2} + 
\frac{\hat{f}'^{2}}{4}
\right)
\stackrel{(\mathcal H)}{\sim} 
-
\left[
\left( \frac{ \omega^{2}}{c_{t}c_{r}} \right)
\,  \frac{1}{ x^{p} }
+
\frac{1}{4} \left( \frac{p}{2}\right)^{2} \frac{1}{x^{2}}
\right]
\; ,
\label{eq:eff-potential_off-shell}
 \end{equation}
 which involves the effective coupling
\begin{equation}
 \lambda_{\rm eff} = \Theta_{\rm eff}^{2}
 =
 \frac{\omega^{2} }{ c_{t} c_{r} }
 \; ,
\label{eq:nh-potential-coupling_off-shell}
 \end{equation}
 and power-law exponent~(\ref{eq:nh-potential-exponent_off-shell}).
 By comparison with the radial
 derivative terms of the normal-reduced 
equation~(\ref{eq:Klein-Gordon_curved}),
 the scale-invariant case defining
CQM
 occurs {\em if and only if\/}
\begin{equation}
 \left. p \right|_{\rm CQM}
=
\left.
\left( p_{t}+ p_{r} \right)
\right|_{\rm CQM}
= 2
\; ,
\label{eq:CQM-exponents_off-shell}
 \end{equation}
in which case the `extra term'
carries the critical coupling $1/4$ that guarantees the relevant physics in
the strong-coupling regime~\cite{renormalization_CQM}
 (and is absorbed in equation~(\ref{eq:nh-potential-coupling_off-shell})).
 
Furthermore, for our
current purposes, the third term on the right-hand side in
equation~(\ref{eq:Klein-Gordon_curved-0_with-L}),
which stands for the angular-momentum degrees of freedom,
leads to the second term in the second line of equation~(\ref{eq:Klein_Gordon_normal_radial_invariant}).
In turn,
 this implies that the near-horizon
 effective angular-momentum term (to be combined with the reduced
operator $\partial_{r}^{2}$)
is 
 \begin{equation}
 V_{L}
  \stackrel{(\mathcal H)}{\sim} 
  \frac{ B }{    { r_{+}^{2} } } 
\, 
\alpha_{l,D}
  \stackrel{(\mathcal H)}{\sim} 
    \frac{ \alpha_{l,D} }{    c_{r} \,  r_{+}^{2}   } \, 
    x^{- p_{r} }
    \; .
\label{eq:nh-angular-potential_off-shell}
    \end{equation}
    
Notice that, once the CQM condition~(\ref{eq:CQM-exponents_off-shell}) 
is adopted, the dominance of the 
singular term defining CQM
is guaranteed for 
$p_{r} < 2$.
Moreover,
it is the integration with respect to the angular momentum
parameter $\alpha_{l,D}$,
when equation~(\ref{eq:nh-angular-potential_off-shell})
is used for statistical counting,
that produces the
holographic area factor in the 
entropy~(\ref{eq:entropy_angular-radial-components}).

\subsection{Geometrical radial distance}
\label{sec:radial-distance_CT}

From the radial part of the metric,
$d \rho^{2} = B(r) \,  dr^{2}$,
it follows that the invariant radial distance from 
the horizon takes the near-horizon form
\begin{equation}
\rho
  \stackrel{(\mathcal H)}{\sim} 
\int \frac{ 1}{ \sqrt{ c_{r} x^{p_{r} } } }
\, dx 
  \stackrel{(\mathcal H)}{\sim} 
 \frac{ 1}{ \sqrt{ c_{r} }}
\,
\frac{  x^{1 -p_{r}/2 } }{
1 -p_{r}/2 }
\; ,
\label{radial-distance_off-shell}
\end{equation}
 whence the geometric elevation of the `brick wall' becomes
 \begin{equation}
h_{D}
  \stackrel{(\mathcal H)}{\sim} 
 \frac{2}{2 - p_{r} }
 \,
 \frac{ 1}{ \sqrt{ c_{r} }}
\,
 a^{1 -p_{r}/2 }
\label{eq:elevation_off-shell}
\; .
\end{equation}
This generalizes
equation~(\ref{eq:brick_wall_geometrical_elevation_regularized})
and permits the computation of spectral 
functions in terms of a geometric,
coordinate-invariant quantity.
Notice that this procedure breaks down for $p_{r} \geq 2$,
with the limiting case $p_{r}=2$ leading to a logarithmic integration.
As a result, 
\begin{equation}
p_{r} < 2
\; 
\label{eq:p-r_upper-bound}
\end{equation}
is a necessary condition 
for consistency of the framework.
The logarithmic limiting case $p_{r}= 2$ will be discussed elsewhere.

\subsection{Hawking temperature and conical singularity}
\label{sec:Hawking-temperature_CT}

  The two-dimensional Euclidean metric  
$ \left.
 ds^{2}
\right|_{ d \Omega = 0 }
$
(restricted to the time-radial sector)
admits the near-horizon approximation
  \begin{eqnarray}
 \left.
 ds^{2}
\right|_{ d \Omega = 0 }
  & = &
   A \, dt_{E}^{2}
  + 
B \, dr^{2}
  \; ,
  \\
   &   \stackrel{(\mathcal H)}{\sim}  &
    c_{t} 
    \,
    x^{p_{t}}
    \,
    dt_{E}^{2}
   + 
   \frac{ 1 }{ c_{r} } 
   \,  x^{-p_{r}} 
   \,  dx^{2}
    \\
    &   \stackrel{(\mathcal H)}{\sim}  &
    \left(
\frac{ 2 - p_{r} }{2}
\right)^{ \frac{ 2 p_{t} }{ 2 - p_{r} } }
    c_{t} 
\,
   \left( c_{r}  \right)^{ \frac{  p_{t} }{ 2 - p_{r} } }
   \,
    \rho^{ \frac{ 2 p_{t} }{ 2 - p_{r} } }
    \,
        dt_{E}^{2}
      +
      d \rho^{2}
  \; ,
  \label{eq:nh-Euclidean}
    \end{eqnarray}
 which takes a conical form
$
 \left.
 ds^{2}
\right|_{ d \Omega = 0 }
=
d \rho^{2} + 
 \alpha^{2} 
\rho^{2} dt_{E}^{2}
$,
ultimately leading to the Hawking temperature,
 {\em if an only if\/}
  \begin{equation}
  \left.
p \right|_{\rm thermal}
=
\left.
\left( p_{t}+ p_{r} \right)
\right|_{\rm thermal}
= 2
  \; .
\label{eq:Hawking-exponents_off-shell}
  \end{equation}
Moreover, the factor $\alpha$
in the Euclidean-time part 
of the metric,
which defines the conical angular deficit\footnote{This 
parameter $\alpha$ is, of course, unrelated to the angular-momentum coupling~(\ref{eq:ang-momentum_coupling}).},
becomes 
$\alpha = \left( p_{t} \sqrt{ c_{r} c_{t} } \right)/ 2$.
In particular, 
removal of the conical singularity 
with a periodic time involves a rescaled 
angular-type coordinate $\chi$ 
of period $2 \pi$,
such that
$
d \chi
=     
\alpha
\,
    dt_{E}
 $;  
correspondingly,
the inverse Hawking temperature becomes
 \begin{equation}
 \beta
 =
 \frac{ 4 \pi }{ p_{t}  }
 \,
 \sqrt{
 \frac{1}{
 c_{t} \, c_{r}
 } }
 \; .
 \end{equation}
Interestingly, the condition~(\ref{eq:Hawking-exponents_off-shell})
is identical to 
conformality, equation~(\ref{eq:CQM-exponents_off-shell}),
within a generalized class of metrics.
This verifies one of the
attributes of the concept of {\em conformal tightness\/}
spelled out in the Introduction.

\subsection{Surface gravity}
\label{sec:surface-gravity_CT}

From the general definition~(\ref{eq:surface_gravity_general}),
based on  the timelike Killing vector
$\mbox{\boldmath  $\xi$ }\!    = \partial_{t}$,
the surface gravity can be computed
geometrically from the limiting 
form of the metric coefficients
at the horizon. 
In particular, for any metric diagonalized in the 
$(t,r)$
sector,
the relation
\begin{equation}
\kappa
 \stackrel{(\mathcal H)}{=} 
\sqrt{
-
g^{tt}  \, g^{rr} \,
\left(
\frac{g_{tt,r} }{2}
\right)^{2}
}
\label{eq:surface_gravity_diagonal-t-r-metrics}
\; 
\end{equation}
follows by straightforward algebra.
In addition,
for the family of metrics~(\ref{eq:off-shell_metric}),
\begin{equation}
\kappa
 \stackrel{(\cal H)}{=}
\sqrt{
\frac{ \left( A' \right)^{2} }{ 4 AB}
}
 \stackrel{(\cal H)}{\sim}
\tilde{\kappa}
\;
x^{[(p_{t}+ p_{r})-2]/2}
\; ,
\label{eq:surface-gravity_off-shell}
\end{equation}
with
\begin{equation}
\tilde{\kappa}
=
\frac{ p_{t}
\,
\sqrt{ c_{r} c_{t}  }
}{2}
\; ,
\label{eq:surface-gravity_off-shell-parameter}
\end{equation}
whence
$\kappa$
{\em has a finite and nonzero value\/}
provided that the necessary condition 
\begin{equation}
\left.
\left( p_{t}+ p_{r} \right)
\right|_{\rm gravity}
 = 2
\end{equation}
be satisfied.
This condition is
identical to
equation~(\ref{eq:Hawking-exponents_off-shell})
for consistency of the Hawking temperature,
i.e.,
 equivalent to the removal of the conical singularity.
The argument above verifies that $\kappa$ is inextricably linked to
the Hawking temperature,
viz., the expected 
identification~(\ref{eq:temperature-gravity_connection}) 
is indeed maintained.

\subsection{Spectral density and entropy}
\label{sec:spectral-density_CT}

The spectral function that generalizes
equation~(\ref{eq:WKB_number_of_states_2})
is computed in a similar manner
by any of the techniques mentioned in section~\ref{sec:area_law},
and further developed in~\ref{sec:spectral-integral}.
This requires the use of the 
effective potentials $V_{\rm eff}$
and $V_{L}$
given by Eqs.~(\ref{eq:eff-potential_off-shell})
and
(\ref{eq:nh-angular-potential_off-shell}).
Notice that, if $p \neq 2$,
the radial measure in
equation~(\ref{eq:WKB_number_of_states_aux-comp_generalized})
fails to be the scale invariant $dx/x$ that ultimately led to the 
trivial `extra scaling factor'~(\ref{eq:angular-radial_entropy-resolution}), 
with a holographic entropy.
By contrast, 
the ensuing modified spectral function and entropy
derived from equation~(\ref{eq:WKB_number_of_states_aux-comp_generalized})
have additional scale factors associated with
the  horizon curvature and radius (which are large scales compared to the
Planck length)---these generate the breakdown of
the trivial scaling~(\ref{eq:angular-radial_entropy-resolution}),
 along with the area law for the entropy.
This can be seen by performing
the integrals in
equation~(\ref{eq:WKB_number_of_states_aux-comp_generalized});
thus, 
from equation~(\ref{eq:WKB_number_of_states_dimensionless_generalized}),
the scaling of the spectral function for $p>2$
takes the form
\begin{equation}
    N (\omega)
\stackrel{(\mathcal H)}{\propto}
\left(
 \sqrt{ \lambda_{\rm eff} }
\right)^{D-1} 
\,
c_{r}^{D/2 -1}
    \,
a^{ - (q-1)}
\;
{\mathcal A}_{D-2}
  \; .
\label{eq:spectral-number_off-shell}
\end{equation}
As shown in~\ref{sec:spectral-integral},
the additional exponent 
\begin{equation}
q = (D-1) \, \frac{ p_{t} }{ 2} + \frac{p_{r}}{2}
\; 
\label{eq:modified_q-exponent}
\end{equation}
[i.e.,
$q = (D/2-1) \,  p_{t} +p/2$]
ultimately generates the breakdown of the
 scaling relations 
and
 holographic scaling,
unless a precise combination of $p_{t}$ and
$p_{r}$ conspires to restore the conformal theory.
Once the replacement 
of the coordinate parameter $a$ by the geometric elevation is
made via
equation~(\ref{eq:elevation_off-shell}),
the scaling of the spectral function becomes 
\begin{equation}
 N(\omega)
\simeq
\biggl(
\frac{\omega}{  2 \tilde{\kappa} } \biggr)^{D-1}
\;
\left(
\frac{
{\mathcal A}_{D-2} }{ \left[ h_{D} \right]^{D-2} }
\right)
\;
\times
\left[
\left( h_{D} \right)^{2}
\,
c_{r}
\right]^{ \frac{(1-q)}{(2-p_{r})} + \frac{D-2}{2} }
\; ,
\label{eq:WKB_number_of_states_scaled_generalized}
\end{equation}
where
$\tilde{\kappa}$
is the auxiliary parameter
of equation~(\ref{eq:surface-gravity_off-shell-parameter}).
The last factor in equation~(\ref{eq:WKB_number_of_states_scaled_generalized}),
\begin{equation}
\left.
\left(
\begin{array}{c}
{\rm `Extra}
\\
{\rm Scaling}
\\
{\rm Factor'}
\end{array}
\right)
\right|_{ \beta \equiv \beta_{H} }
\simeq
\left[
\left( h_{D} \right)^{2}
\,
c_{r}
\right]^{ \frac{(1-q)}{(2-p_{r})} + \frac{D-2}{2} }
\; ,
\label{eq:extra-scaling_generalized}
\end{equation}
is transferred from the spectral function to the entropy
via the algorithm~(\ref{eq:entropy_formula}).
In principle, such dependence would modify
the scaling because it depends on ratios of 
the size of curvature and horizon
radius to the Planck length.
Then, the removal of this spurious
scaling~(\ref{eq:extra-scaling_generalized})
is guaranteed if and only if 
\begin{equation}
\frac{ (1-q) }{ (2-p_{r}) } + \frac{(D-2)}{2} = 0
\; ;
\end{equation}
when combined with the definition of $q$,
equation~(\ref{eq:modified_q-exponent}),
a linear relation between 
$p_{t}$
and
$p_{r}$
is obtained, which coincides with
the CQM condition~(\ref{eq:CQM-exponents_off-shell}).
The argument above proves that: 
\begin{quotation}
\noindent
{\em Any deviation from conformal
quantum mechanics would lead to a breakdown of the 
Bekenstein-Hawking area law for the entropy.}
\end{quotation}
Moreover, it should be noticed that,
as the dimensions of $c_{t}$ and $c_{r}$
depend on the chosen exponents $p_{t}$ and $p_{r}$,
with $[\tilde{\kappa}] = \left[ {\rm length} \right]^{- p/2}$,
this would introduce yet additional scaling modifications
via the first factor in 
equation~(\ref{eq:WKB_number_of_states_scaled_generalized}).
Thus,
at an even deeper level,
the thermodynamics is more tightly constrained by the fact
that 
the parameter
$\tilde{\kappa} $ is not a genuine surface gravity---simply put,
neither the surface gravity nor the temperature
can be consistently defined, as discussed in
subsections~\ref{sec:Hawking-temperature_CT}
and \ref{sec:surface-gravity_CT}.
The case $p<2$ leads to a non-homogeneous spectral function $N(\omega)$, and the scaling relations above are further 
disrupted.
In short,
once the restriction~(\ref{eq:Hawking-exponents_off-shell})
is enforced for consistency,
the conformal theory applies with $p=2$,
but with a possibly modified 
angular-momentum term as in 
equation~(\ref{eq:nh-angular-potential_off-shell}).
In that theory,
the 
scaling~(\ref{eq:extra-scaling_generalized})
becomes trivial and holography is restored.
In effect,
regardless
of the value of 
$p_{r} <2$ in 
equation~(\ref{eq:nh-angular-potential_off-shell}),
the basic entropy expression~(\ref{eq:entropy_brick_wall})
will still hold true---possibly with a modified numerical prefactor that
carries  the value
$q = 1+ (D/2 -1) p_{t}$
(instead of $D/2$).
This concludes the proof of the conformal nature of the 
area law for the entropy.

\subsection{Further characterization and examples of conformal tightness}
\label{sec:further-CT_examples}

The results of the previous subsections
give a basic characterization of the concept of
 {\em conformal tightness\/}
spelled out in the Introduction, including both the Hawking-temperature 
identification
and the preservation of holographic scaling.

Let us now sum up the various 
constraining conditions 
on the values of $p_{t}$ and $p_{r}$:
a horizon restriction~(\ref{eq:p-r_horizon-condition}),
a thermal restriction~(\ref{eq:Hawking-exponents_off-shell}),
and the geometric distance and CQM vs.\ angular-momentum 
restriction~(\ref{eq:p-r_upper-bound}).
When the near-horizon expansions are
analytic, with integer exponents, the set of restrictions
uniquely selects the original case of CQM,
 with $p_{t}=p_{r} =1$.
 The corresponding metric is 
a modification of equation~(\ref{eq:RN_metric}), 
with a similar analytic structure,
\begin{equation}
ds^{2}
=
- C(r) f(r) dt^2 + D(r)  
[f(r)]^{-1} dr^2 + r^{2} 
\, d \Omega^{2}_{(D-2)}
\;  ,
\label{eq:generalized_Schwarzs_metric}
\end{equation}
and supplemented with the restrictions: firstly that
 \begin{equation}
f(r)
\stackrel{(\mathcal H)}{\sim} 
f'_{+} x
\; 
\label{eq:scale-factor_nonextremal}
\end{equation}
(with $f'_{+} \neq 0$
defining a nonextremal black hole solution);
and secondly,
that
both $C$ and $D$ should tend to constant, nonzero values 
\begin{equation}
C \stackrel{(\mathcal H)}{\sim} C_{*}
\; \; \; \; \; \; \; \;
D \stackrel{(\mathcal H)}{\sim} D_{*}
\; 
\label{eq:metric-prefactors}
\end{equation}
at the horizon.
Notice that
the form of the metric~(\ref{eq:generalized_Schwarzs_metric}) can be recast into our original 
parametrization of Eqs.~(\ref{eq:off-shell_metric})
and (\ref{eq:GM_scale-factor}) 
via
$\hat{f}^{2} \equiv A/B = f^{2} \, C/D$.

Alternatively,
by redefining the radial coordinate
via $r^2 \rightarrow r^2 D(r)$,
this generalized family takes the form
\begin{equation}
ds^{2}
=
- {C}(r) f(r) dt^2 + {D}(r)  \left\{ 
[f(r)]^{-1} dr^2 + r^{2} 
\, d \Omega^{2}_{(D-2)}
\right\}
\;  ,
\label{eq:generalized_Schwarzs_metric2}
\end{equation}
with the conditions~(\ref{eq:scale-factor_nonextremal}) and
(\ref{eq:metric-prefactors})
to be enforced just as before.
It should be noticed that
$r$ is no longer the area coordinate
in equation~(\ref{eq:generalized_Schwarzs_metric2})
[except in the case when 
${D}(r)  = 1$]; 
as customary,
we have kept the same symbol $r$, even though
the new $r$ is a different coordinate.

In short, as shown
throughout this section,
{\em the modified metrics
do not change the essence of the
conformal character or the fundamental
 thermodynamic results.\/}
As a simple illustration of this generalization, let us consider
the stringy solution
corresponding to a five-dimensional extended-supergravity
and non-extremal 
metric~\cite{stringy_metric,ortin}
\begin{equation}
ds^{2} =
- \left( 
1 -\frac{r_{0}^{2} }{ r^{2}} 
\right) \, 
F^{-2/3} dt^2 
+ 
F^{1/3} \, 
\left[ 
\left(
1 - \frac{r_{0}^{2} }{ r^{2}} 
\right)^{-1} 
dr^2 + r^2 d \Omega_{(3)}^2 
\right]
\; ,
\label{eq:susy_metric}
\end{equation}
which is not of the form~(\ref{eq:RN_metric}), but 
does fall into the class~(\ref{eq:generalized_Schwarzs_metric2}),
with the obvious substitutions
$f = 1 - r_{0}^{2}/r^{2}$,
${C} = F^{-2/3}$,
and 
${D} =  F^{1/3}$
[where $F(r) $ 
has a finite
value dependent on the free
parameters related to the three charges and mass].
This model can be further extended to $D$ dimensions~\cite{ortin},
\begin{equation}
\! \! 
\! \! \! \! \! \! \! \! \! \! \! \! 
\! \! \! \! \! \! \! \! \! \! \! \! 
\! \! \! \! \! \! \! \! \! \! \! \!
ds^{2} =
- \left( 
1 -\frac{r_{0}^{D-3} }{ r^{D-3}} 
\right) \, F^{-(D-3)/(D-2)} 
dt^2 
+ 
F^{1/(D-2)} 
 \, \left[ \left(
1 - \frac{r_{0}^{D-3} }{ r^{D-3}} 
\right)^{-1} 
dr^2 + r^2 d \Omega_{(D-2)}^2 
\right]
,
\,
\end{equation}
following the same pattern~(\ref{eq:generalized_Schwarzs_metric2});
and similar considerations could be applied to other geometries.
The results derived in this paper
not only exhibit the conformal tightness property of the metric
but also provide a simple computational framework
for the evaluation of all the relevant thermodynamic quantities.

\section{Outstanding issues and conclusions}
\label{conclusions}

Our analysis highlights the fundamental nature of the SO(2,1) conformal
symmetry for black hole thermodynamics.
This is substantiated by
comprehensive analytical and scaling arguments  using a scalar field probe.
In essence, at first, the angular degrees of freedom appear to 
yield the Bekenstein-Hawking area scaling~(\ref{eq:holo_entropy})
for the scalar-field contribution to the entropy. 
However, the apparent holographic scaling
is preserved 
{\em if and only if\/} the near-horizon physics is conformal,
as revealed by the radial degree of freedom.
This feature of the thermodynamics 
is tested by extending the class of metrics
and showing the ensuing {\em conformal tightness\/}.
In this manner, our work extends the scope of~\cite{BH_thermo_CQM,semiclassical_BH_thermo}
and \cite{holographic_scaling_I}.

The program
can be further expanded in the followings ways:
considering other, non-scalar, fields;
verifying the conformal tightness
of holographic scaling for higher orders of the 
heat kernel expansion;
extending  the class of metrics  of subsection~\ref{sec:further-CT_examples}
to axisymmetric spacetimes;
and finding a deeper physical interpretation of this intriguing conformal symmetry.
These extensions, which will
 be discussed elsewhere, are technically relevant. In effect, any violations of the area law for particular fields would render the holographic result unlikely. In this sense, the verification of this property for a scalar probe is a nontrivial first step.

Given the scope of the conformal
framework for black hole thermodynamics, 
our arguments suggest 
the need for a deeper characterization
of this intriguing property---a topic that
will be expanded in a forthcoming paper.
This
exploration would
entail unearthing the geometrical meaning of the symmetry
{\em vis-\`{a}-vis\/}
the near-horizon approximation
and possibly establishing a connection with other frameworks,
e.g., the approach
of Refs.~\cite{carlip:near_horizon,solodukhin:99}.

\ack

We are grateful to the referees for their insightful comments.
One of us (CRO) would like to thank 
the Kavli Institute for Theoretical Physics for generous support and for 
providing a wonderful environment during the initial stages of this work. 
Fruitful conversations with 
Don Marolf
are also acknowledged. 
This work was partially supported
by the National Science Foundation under Grant
Nos 0602340  (HEC)
and  0602301 and PHY05-51164 (CRO), and
by the University of San Francisco Faculty Development Fund
(HEC).

\appendix

\section{Semiclassical spectral analysis:
evaluation 
of the `near-horizon spectral integral'
for black hole thermodynamics}
\label{sec:spectral-integral}

The spectral function  $N (\omega) $
is evaluated 
via a semiclassical 
algorithm
\begin{equation}
 N(\omega) 
  = 
\underbrace{
\frac{1}{ \Gamma (D-2) }
\int
d \alpha \,
\alpha^{(D-4)/2} 
}_{\rm angular \; contribution}
\,
 \underbrace{
  \frac{1}{\pi} 
\,
\int
dr
\,
 	k_{ \alpha} (r)
}_{\rm radial \; contribution}
\; ,
\label{eq:semiclassical_number_of_states}
\end{equation}
with the angular summation $\sum_{l,m}$
replaced by a semiclassical integral
and a WKB radial ordinal-number estimate
via a Langer-corrected wavenumber 
 $ k_{\alpha_{l,D}} (r)
 = k_{\alpha_{l,D}} (r_{+}+ x ) 
=
\sqrt{ 
\mathfrak{I}  
-1/ 4 x^{2} }
$,
as mandated by the horizon coordinate singularity.
The 
potential~(\ref{eq:neg-BH-potential})
has two turning points,
$a$ and $x_{1} \equiv x_{1} (\alpha) $.
Correspondingly, in equation~(\ref{eq:semiclassical_number_of_states}),
the WKB algorithm involves the 
turning-point cutoffs 
enforced by the double Heaviside
product 
$\theta  \bigl( x - a \bigr) \, 
\theta  \bigl( x_{1} (\alpha)  - x \bigr)
$;
in addition, a third Heaviside function
enforces the condition $a \leq x_{1}$,
which prevents the inversion of the
turning points $a$ and $x_{1}$,
so that
\begin{eqnarray}
\! \! \! \! \! \! \! \! \! \! \! \!
 N(\omega) 
\stackrel{(\mathcal H)}{\sim}
\frac{ \Theta}{\pi \, \Gamma (D-2) }
\int_{0}^{ \infty} 
& &
d \alpha \,
\alpha^{D/2- 2} 
 \, 
\int_{0}^{ \infty } 
\frac{dx}{x}
\,
\sqrt{ 1 
- 
\frac{ \alpha }{f_{+}' r_{+}^{2} \, \Theta^{2} }
\, x }
\nonumber
\\
& \times &
\theta  \bigl( x - a \bigr) \, 
\theta  \bigl( x_{1} (\alpha ) - x \bigr)
\,
\theta 
\bigl(
\alpha_{\rm max} (a) - \alpha
\bigr)
\; ;
\label{eq:WKB_number_of_states_aux-comp}
\end{eqnarray}
thus,
the `near-horizon
spectral integral'~(\ref{eq:WKB_number_of_states})
is established.

The coordinate-singular nature of the near-horizon 
physics suggests that extra care should be exercised 
when evaluating
the spectral function~(\ref{eq:WKB_number_of_states_aux-comp}).
In this appendix,
we focus on two alternative calculations of this integral, 
while an equivalent phase-space approach follows in~\ref{sec:phase-space}---these 
multiple evaluations confirm the robust nature
of the final outcome.
The physical scales involved in 
equation~(\ref{eq:WKB_number_of_states_aux-comp})
can be explicitly singled out and displayed,
\begin{equation}
 N(\omega) 
\stackrel{(\mathcal H)}{\sim}
\Theta^{D-1} 
\,
\left( f'_{+} \, r_{+}^{2}  \right)^{ (D-2)/2 }
a^{-(D/2-1)}
\;
{\mathcal N}
\; ,
\label{eq:WKB_number_of_states_dimensionless}
\end{equation}
via the introduction of the dimensionless auxiliary
variables shown below.
In equation~(\ref{eq:WKB_number_of_states_dimensionless}),
the dimensionless integral ${\mathcal N}$ can be computed 
 in two different ways:
\begin{enumerate}
\item
As the sequence of a radial plus an angular integration
(in that order),
via
\begin{equation}
\hat{\alpha} 
=
\frac{\alpha}{\alpha_{\rm max} (a)}
= 
\frac{\alpha \,  a}{ f'_{+} \, r_{+}^{2} \Theta^{2} }
\;    ,
\; \; \;  \; \; \;  \; \; \;  \; \; \; 
\zeta =
\hat{\alpha} 
\,
\frac{x}{a}
\; ,
\label{eq:alpha-hat_CQM}
\end{equation}
leading to
\begin{equation}
{\mathcal N}
=
\frac{  1 }{\pi \, \Gamma (D-2) }
\,
\int_{0}^{1} 
d \hat{\alpha}  \, \hat{\alpha}^{D/2-2} 
\,
\int_{\hat{\alpha} }^{1} 
\frac{d \zeta }{\zeta}
\,
\sqrt{ 1 - \zeta } 
\; .
\label{eq:N_radial-plus-angular}
\end{equation}

\item
As the sequence of an 
angular plus a radial integration,
via
\begin{equation}
\hat{\alpha}' = 
\frac{\alpha}{\alpha_{\rm max} (x)}
= 
\frac{\alpha  \, x}{ f'_{+} \, r_{+}^{2} \Theta^{2} }
\;    ,
\; \; \;  \; \; \;  \; \; \;  \; \; \; 
u = \frac{x}{a}
\; ,
\end{equation}
leading to
\begin{equation}
{\mathcal N}
=
\frac{  1 }{\pi \, \Gamma (D-2) }
\,
\int_{1}^{ x_{1}/a }
\frac{d u }{ u^{D/2} }
\,
\int_{0}^{1} 
d \hat{\alpha }'  \, \left( \hat{\alpha  }' \right)^{D/2-2}
\sqrt{ 1  - \hat{\alpha  }' }
\; .
\label{eq:N_angular-plus-radial}
\end{equation}

\end{enumerate}

Either from equation~(\ref{eq:N_radial-plus-angular})
or from equation~(\ref{eq:N_angular-plus-radial}),
the numerical coefficient 
${\mathcal N}$
is found to be
\begin{equation}
 {\mathcal N} 
= 
\frac{  2 \,  B \bigl( D/2 -1, 3/2 \bigr)
}{\pi \, (D-2) \, \Gamma (D-2) }
\; .
\label{eq:spectral-function_coeff}
\end{equation}
Specifically,
equation~(\ref{eq:N_angular-plus-radial})
 yields this result from the angular integral---generating 
a beta function 
$B \bigl( D/2 -1, 3/2 \bigr)$] 
and enforcing ${ x_{1}/a \sim \infty}$.
By contrast,
equation~(\ref{eq:N_radial-plus-angular})
leads to the integral
$\int_{0}^{1} d \hat{\alpha}
\,
\hat{\alpha}^{D/2 -2}
\,
\left[
F(1) - F(\hat{\alpha})
\right]
$, which 
involves the function
\begin{equation}
F (\zeta)
=
\int^{\zeta} 
d \zeta
\,
\frac{\sqrt{ 1 - \zeta }}{\zeta}
=
2 
\sqrt{ 1 - \zeta } 
+
\ln 
\left| 
\frac{ \left( 
\sqrt{ 1 - \zeta } - 1
\right)
}{
\left( 
\sqrt{ 1 - \zeta } + 1
\right)
}
 \right|
\; 
\label{eq:spectral-aux-intergral_CQM}
\end{equation}
and reproduces the same value
after a straightforward integration by parts.

When the generalized 
metrics~(\ref{eq:off-shell_metric})
of section~\ref{sec:conformal_tightness}
are considered,
with effective potentials $V_{\rm eff}$
and $V_{L}$
given by Eqs.~(\ref{eq:eff-potential_off-shell})
and
(\ref{eq:nh-angular-potential_off-shell}),
the spectral integral~(\ref{eq:WKB_number_of_states_aux-comp})
gets modified into 
\begin{eqnarray}
\! \! \! \! \! \! \! \! \! \! \! \!
 N(\omega) 
\stackrel{(\mathcal H)}{\sim}
\frac{  \sqrt{ \lambda_{\rm eff}} }{\pi \, \Gamma (D-2) }
\int_{0}^{ \infty} 
& &
d \alpha \,
\alpha^{D/2- 2} 
 \, 
\int_{0}^{ \infty } 
\frac{dx}{x^{p/2} }
\,
\sqrt{ 1 
- 
\frac{ \alpha }{  c_{r} r_{+}^{2} \, 
 \lambda_{\rm eff} }
\, x^{ p_{t} }
}
\nonumber
\\
& \times &
\theta  \bigl( x - a \bigr) \, 
\theta  \bigl( x_{1} (\alpha ) - x \bigr)
\,
\theta 
\bigl(
\alpha_{\rm max} (a) - \alpha
\bigr)
\; .
\label{eq:WKB_number_of_states_aux-comp_generalized}
\end{eqnarray}
The integrals in
equation~(\ref{eq:WKB_number_of_states_aux-comp_generalized})
can be performed just for the conformal case,
via the generalization of 
Eqs.~(\ref{eq:alpha-hat_CQM})---(\ref{eq:spectral-aux-intergral_CQM});
the outcome of this procedure is
\begin{equation}
 N(\omega) 
\stackrel{(\mathcal H)}{\sim}
\left(
 \sqrt{ \lambda_{\rm eff} }
\right)^{D-1} 
\,
\left( c_{r} \, r_{+}^{2}  \right)^{ (D-2)/2 }
a^{-(q-1)}
\;
{\mathcal N}
\; ,
\label{eq:WKB_number_of_states_dimensionless_generalized}
\end{equation}
where the modified exponent is
$
q = (D-1) \,  p_{t}/ 2 + p_{r}/ 2
$ 
(equation~(\ref{eq:modified_q-exponent})),
which is seen to be different from $D/2$
and spoils all the scaling relations.
In addition, the numerical coefficient---which replaces
equation~(\ref{eq:spectral-function_coeff})---is
\begin{equation}
{\mathcal N} 
= 
\frac{   B \bigl( D/2 -1, 3/2 \bigr)}
{\pi \, (q-1) \, \Gamma (D-2) }
\; .
\end{equation}
Incidentally, 
in equation~(\ref{eq:WKB_number_of_states_aux-comp_generalized}), 
we ignored the extra term proportional to $1/x^2$ in the effective potential~(\ref{eq:eff-potential_off-shell}).
This term only matters for $p=2$; otherwise, if $p>2$,
it merely generates a correction scaled by $a^{p-2}/\lambda_{\rm eff}$,
which is of the order of the very small ratio $\left( a/r_{+} \right)^{p-2}$ in the radicand, and can be safely neglected.
If $p<2$, then the correction is more complex as the spectral function is no longer homogeneoeus---and 
the scaling relations are
definitely altered in a nontrivial manner.

\section{Phase-space arguments}
\label{sec:phase-space}

The phase-space method
for the semiclassical statistical counting of states provides further
insight into our conformal-tightness arguments.
In this method, the 
reduced Klein-Gordon 
equation
involves a simple Hamiltonian formulation
(with
effective Hamiltonian
$\mathcal{H}_{\rm eff}$),
leading to the 
spectral function
\begin{equation}
N (\omega)
\approx
\int 
d^{d} x
\,
\int
\frac{ d^{d} p}{ (2 \pi )^{d} }
\;
\theta (- {\mathcal H_{\rm eff}})
\; ,
\end{equation}
in $d=D-1$
spatial dimensions.
In a straightforward
 multidimensional approach,
the spectral function is written
as a configuration-space integral
and 
{\em all\/} the generalized momenta are first integrated
out simultaneously.
The ensuing  integral becomes
\begin{equation}
N ( {\omega})
\approx
\frac{ \Omega_{(d-1)}}{ d \, ( 2 \pi)^{d} }
\,
\int 
 d V_{(d)} 
\,
\,
\| \vec{k}  (\vec{x}) \|^{d}
\; ,
\label{eq:phase_space_spectral_number_explicit_fully_multidim}
\end{equation}
where
$ d V_{(d)} 
= d^{d} x \, 
\sqrt{\gamma}$
is the $d$-dimensional spatial volume element,
$\gamma$ is the spatial metric,
and 
$\| \vec{k}  (\vec{x}) \|$
is defined from
the generalized WKB framework.
In the case of spherical symmetry,
\begin{equation}
N ( {\omega})
\approx
\frac{ [\Omega_{(d-1)}]^{2} }{ d \, ( 2 \pi)^{d} }
\,
\int 
dr 
\,
 [ \gamma_{rr} ]^{-(d-1)/2}
\,
r^{d-1} 
\,
\left[
\tilde{k}(r)
\right]^{d}
\; ,
\label{eq:phase_space_spectral_number_polar_coords_explicit_alternative}
\end{equation}
where
$
\tilde{k}( r ) 
= [ \gamma_{rr} ]^{1/2} 
\, 
\| \vec{k}  (r) \|
$
is the covariant-component form of the radial counterpart of 
$\| \vec{k}  (\vec{x}) \|$.
Details can be found in Ref.~\cite{semiclassical_BH_thermo}.

For the computation
of the spectral functions in our
paper,
all that is needed
is the reversal to $D= d+1$ and the replacements:
\begin{itemize}
\item
\begin{equation}
\tilde{k}( r ) 
\stackrel{(\mathcal H)}{\sim}
k(x) \equiv
\sqrt{-V_{\rm eff}(x)}
=
 \frac{ \sqrt{ \lambda_{\rm eff}} }{ x^{p/2} }
\; ,
\end{equation}
where $V_{\rm eff}(r)$ is given in 
equation~(\ref{eq:eff-potential_off-shell}).

\item
\begin{equation}
\gamma_{rr}
\stackrel{(\mathcal H)}{\sim}
\frac{1}{c_{r} x^{ p_{r} }}
\; .
\end{equation}
\end{itemize}
As a result, the 
competition between the angular and radial degrees
of freedom acquires its factorized form because
of the 
 corresponding multiplicative
structure in phase space; specifically,
\begin{equation}
\! \! \! \! \! \! \! \! \! \! \! \!
 \! \! \! \! \! \! \! \! \! \! \! \!
 N(\omega) 
  \stackrel{(\mathcal H)}{\sim} 
\frac{ \Omega_{(d-1)} }{ d \, ( 2 \pi)^{d} }
\,
\underbrace{
\Omega_{(d-1)} 
r_{+}^{D-2} }_{\rm angular \; piece \; = \; {\mathcal A} } 
\,
\int 
dx 
\,
\underbrace{
\left[ k(x) \right]^{D-1}
}_{\rm radial \; interaction}
\,
\underbrace{
\left[
c_{r} x^{ p_{r}}  
\right]^{(D-2)/2}
}_{\rm metric  \; factors}
\;  .
\label{eq:INTRO_phase_space_spectral_number_polar_coords_explicit_QFT_n-h}
\end{equation}
This integral expression reduces to the conformal 
spectral function~(\ref{eq:WKB_number_of_states_2})
for $p=2$ and to the generalized case,
equation~(\ref{eq:WKB_number_of_states_dimensionless_generalized}),
for arbitrary values of $p$ and $p_{r}$.

Incidentally, even though the
derivation and use of 
equation~(\ref{eq:INTRO_phase_space_spectral_number_polar_coords_explicit_QFT_n-h})
may appear to be more involved,
the phase-space method reveals even more clearly
the separation of the degrees of freedom 
directly arising from the angular and radial variables.

\end{document}